\documentclass[conference]{IEEEtran}
\usepackage{cite}

\ifCLASSINFOpdf
\usepackage[pdftex]{graphicx}
\else
\fi
\usepackage{amsmath}
\usepackage{amsfonts} 
\usepackage{amsthm}
\usepackage{amssymb}
\usepackage{amsthm} 
\usepackage{mathtools} 
\usepackage{textcomp} 
\usepackage{xcolor}
\usepackage{clrscode}

\newtheorem{algorithm}{Algorithm}
\newtheorem{CodeListing}{Code Listing}

\usepackage{BOONDOX-ds}

\usepackage{bbm}

\newcommand{\pfun}{\mathop{\hbox{$\to$\kern-7pt\raise.9pt\hbox{\scalebox{1}[.55]{$|$}}\kern4pt} }}

\usepackage{array}

\begin{document}

\title{Easy Acceleration with Distributed Arrays
}

\author{\IEEEauthorblockN{
Jeremy Kepner, Chansup Byun, LaToya Anderson, William Arcand, David Bestor, William Bergeron, \\ Alex Bonn, Daniel Burrill, Vijay Gadepally, Ryan Haney, Michael Houle, Matthew Hubbell, Hayden Jananthan, \\ Michael Jones, Piotr Luszczek, Lauren Milechin, Guillermo Morales, Julie Mullen, \\ Andrew Prout,  Albert Reuther, Antonio Rosa, Charles Yee, Peter Michaleas
\\
\IEEEauthorblockA{
MIT
}}}
\maketitle
\begin{abstract}
High level programming languages and GPU accelerators are powerful enablers for a wide range of applications.  Achieving scalable vertical (within a compute node), horizontal (across compute nodes), and temporal (over different generations of hardware) performance while retaining productivity requires effective abstractions.   Distributed arrays are one such abstraction that enables high level programming to achieve highly scalable performance.  Distributed arrays achieve this performance by deriving parallelism from data locality, which naturally leads to high memory bandwidth efficiency.  This paper explores distributed array performance using the STREAM memory bandwidth benchmark on a variety of hardware.  Scalable performance is demonstrated within and across CPU cores, CPU nodes, and GPU nodes.  Horizontal scaling across multiple nodes was linear.  The hardware used spans decades and allows a direct comparison of hardware improvements for memory bandwidth over this time range; showing a 10x increase in CPU core bandwidth over 20 years, 100x increase in CPU node bandwidth over 20 years, and 5x increase in GPU node bandwidth over 5 years.  Running on hundreds of MIT SuperCloud nodes simultaneously achieved a sustained bandwidth $>$1 PB/s.  
\end{abstract}

\begin{IEEEkeywords}
GPUs, distributed arrays, parallel processing, vertical scaling, horizontal scaling
\end{IEEEkeywords}

%
\IEEEpeerreviewmaketitle

\section{Introduction}
\let\thefootnote\relax\footnotetext{Distribution Statement A.  Approved for public release.  Distribution is unlimited.  This material is based upon work supported by the Under Secretary of Defense for Research and Engineering under Air Force Contract No. FA8702-15-D-0001 or FA8702-25-D-B002. Any opinions, findings, conclusions or recommendations expressed in this material are those of the author(s) and do not necessarily reflect the views of the Under Secretary of Defense for Research and Engineering.
Use of this work is controlled by the human-to-human license listed in Exhibit 3 of https://doi.org/10.48550/arXiv.2306.09267
}


Easy-to-use high level programming languages (e.g., Matlab \cite{higham2016matlab}, Octave \cite{eaton2012gnu}, Python \cite{van1995python}, and Julia \cite{bezanson2017julia},) enable millions of programmers and play an important role in computational science and high performance computing.  Accelerators, most notably GPUs (Graphics Processing Units), have become key enablers for a wide range of computationally intensive workloads, particularly in AI (Artificial Intelligence) \cite{reuther2022ai, reuther2023lincoln}.  The capability of CPUs (Central Processing Units) continues to increase as the number of processing cores on a CPU grows.  The number of distinct compute nodes in supercomputers and cloud systems ranges from thousands to millions \cite{pilz2025trends}.  A key challenge is maintaining the ease-of-use of high level programming languages while achieving scalable performance across different dimensions: vertical (within a compute node), horizontal (across compute nodes), and temporal (over different generations of hardware).

Vertical scaling within nodes often relies on threading approaches (e.g., OpenMP \cite{dagum1998openmp}, pthreads \cite{nichols1996pthreads}, and Cilk \cite{blumofe1995cilk}) that leverage shared memory model.  Threading allows  libraries (e.g., MKL \cite{wang2014intel}, MAGMA\cite{agullo2009numerical}, SPIRAL\cite{franchetti2018spiral}, and GraphBLAS\cite{kepner2016mathematical, davis2019algorithm}) written in low-level languages (e.g., C\cite{kernighan1988c}, C++\cite{stroustrup2013c++}, and Fortran\cite{press1992numerical}) to deliver parallel performance at the function level that can be readily called by a high level programming language.  Vertical scaling on accelerators often follows a similar approach leverage technologies that are designed for more heterogenous accelerator architectures  (e.g., CUDA\cite{sanders2010cuda} and OpenCL\cite{munshi2011opencl}).

Horizontal scaling across nodes often relies on messaging passing (e.g., MPI\cite{gropp1999using} and SHMEM\cite{chapman2010introducing}) approaches that can work with both shared and distributed memory.  Although more challenging than threading, library writers can create libraries (e.g., PETSc\cite{balay2019petsc}, Trilinos\cite{heroux2005overview}, and VSIPL++\cite{lebak2005parallel}) in low level languages that horizontally scale to thousands of nodes.  Message passing and threading approaches are often used together with threading used for vertical scaling and message passing used for horizontal scaling.   In the common case where the nodes only need to communicate with a central coordinator, then simpler client-server \cite{sinha1992client} or map-reduce \cite{dean2008mapreduce, byun2016llmapreduce} approaches are often used for horizontal scaling.

Distributed arrays (also referred to a PGAS -- partitioned global address spaces\cite{yelick2007productivity}) are a complimentary approach that can be used for both horizontal and vertical scaling, often leveraging threading and message passing as underlying enabling technologies.  Distributed arrays can be implemented effectively in low level languages \cite{de2015partitioned, amarasinghe2023compiler} and high level languages (e.g., Matlab\cite{choy2004star, moler2020history}, Octave\cite{Kepner2009}, Python\cite{byun2023ppython, shajii2023codon}) using both shared and distributed memory hardware.  Many message passing libraries present themselves to users as distributed arrays \cite{balay2019petsc, heroux2005overview, lebak2005parallel}.  Distributed arrays align well with high level languages as large arrays are often a core concept within high level languages and operating on large arrays as a whole (vectorization \cite{cai2005performance, birkbeck2007dimension}) is an important optimization technique.

This paper presents a short case study on the performance of distributed arrays in high level languages using the STREAM (hereafter ``Stream'') memory bandwidth benchmark \cite{McCalpin2005}.  Users of high level languages often use memory extensively to simplify programming which places significant pressure on memory bandwidth.  Over the decades, our team has found that Stream performance is a good indicator of the real user performance in high level languages.  The benchmark is implemented in Matlab, Octave, and Python using the pMatlab\cite{Kepner2009} and pPython\cite{byun2023ppython} distributed array libraries.  Performance is measured on hundreds of nodes of varying CPU and GPU hardware spanning multiple decades.  The outline of the rest of the paper is as follows.  First, a description of various parallel programming models is provided with an emphasis on the distributed arrays programming model.  Next, the Stream benchmark is presented along with a corresponding distributed array parallel design.  Subsequent sections describe the code, benchmarking hardware, and performance results followed by the conclusion and a discussion of potential future work.

\section{Parallel programming models}

  The client-server and map-reduce models are often the simplest parallel programming model.  These models are used when a problem can be broken up into a set of completely independent tasks that the workers (clients) can process without communicating with each other. The central constraint of the client-server and map-reduce models are that each worker communicates only with the leader (server) and requires no knowledge about what the other workers are doing.  This constraint is very powerful and is enormously simplifying.  Furthermore, these models have proven themselves very robust and many parallel programs follow this model.

The message passing model is in many respects the opposite of the client-server and map-reduce models.  The message passing model requires that any process (denoted by a $P_{ID}$) be able to send and receive messages from any other $P_{ID}$. The infrastructure of the message passing model is fairly simple. This infrastructure is most typically instantiated in the parallel computing community via the Message Passing Interface (MPI) standard \cite{gropp1999using}. The message passing model requires that each processor have a unique identifier ($P_{ID}$) and know how many other processors ($N_P$) are working together on a problem (in MPI terminology $P_{ID}$ and $N_P$ are referred to as the processor ``rank'' and the  ``size'' of the MPI world).  Any parallel program can be implemented using the message passing model. The primary drawback of this model is that the programmer must manage every individual message in the system; therefore, the model often requires a great deal of additional code and may be extremely difficult to debug. Nevertheless, there are certain parallel programs that can be implemented only with a message passing model.

\begin{figure}[t!]
\centerline{\includegraphics[width=\columnwidth]{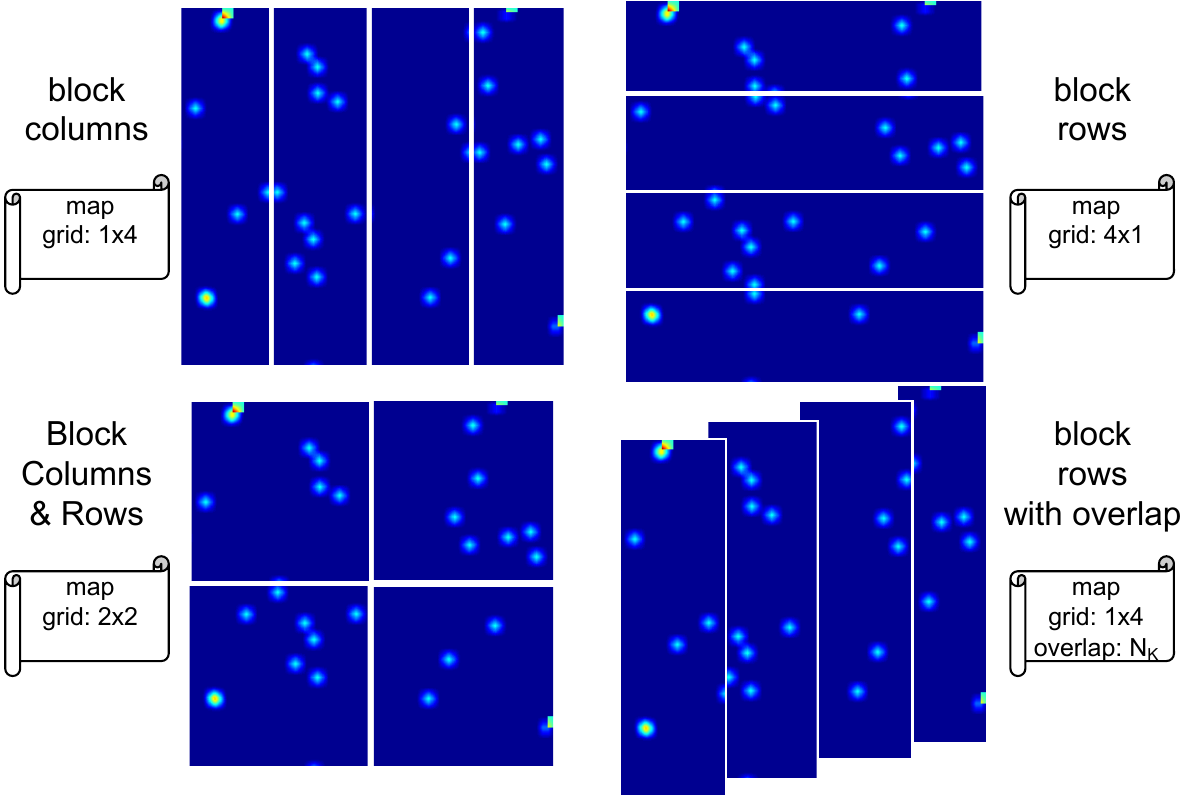}}
\caption{{\bf Distributed Array Mappings} (adapted from \cite{Kepner2009}).
Different parallel mappings of a two-dimensional array.  Arrays can be broken up in any dimension.  A block mapping means that each $P_{ID}$ holds a contiguous piece of the array.  Overlap allows the boundaries of an array
to be stored on two neighboring $P_{ID}$s.}
\label{fig:GlobalArrayMaps}
\end{figure}

  The distributed array model is a compromise between the two models. Distributed arrays impose additional constraints on the program, which allow complex programs to be written relatively simply. In many respects, it is the
most natural parallel programming model for high level languages because it is implemented using arrays, which are often the core data type in a high level language. Briefly, the distributed array model creates distributed arrays in which each $P_{ID}$ stores or owns a piece of the whole array.  Additional information is stored in the array so that every $P_{ID}$ knows which parts of the array the other $P_{ID}$s have. How the arrays are broken up among the $P_{ID}$s is specified by a map~\cite{lebak2005parallel, Kepner2009, byun2023ppython}.  For example, Figure~\ref{fig:GlobalArrayMaps} shows a matrix broken up by rows, columns, rows and columns, and columns with some overlap.  The concept of breaking up arrays in different ways is one of the key ideas in parallel computing.

  Computations on distributed matrices are usually performed using the ``owner computes'' rule, which means that each
$P_{ID}$ is responsible for doing a computation on the data it is storing locally. Maps can become quite complex and express virtually arbitrary distributions.  One example of this complexity is the case in which
a boundary of an array is required by more than one $P_{ID}$ and will be implicitly communicated to complete the computation.  Another example are pipelines which can be implemented by mapping different arrays to different sets of  $P_{ID}$s.

 \section{STREAM}

  The Stream benchmark \cite{McCalpin2005} is a simple program that is very useful for illustrating several important parallel programming concepts.  Stream is representative of a wide class of pleasingly parallel memory intensive programs.  Stream highlights the importance of data locality and how to extract concurrency from data locality. Stream also shows performance implications of different data-parallel access approaches.  Finally, Stream sheds light on the performance trade-offs of multicore processors.

Stream creates several large $N$ element vectors and repeatedly performs basic operations on these vectors.  Because the number of operations performed on each element of a vector is small, Stream performance is limited by the bandwidth to main memory.  Stream provides an excellent illustration of the performance trade-offs of multicore processors that all share the same access to memory.

  The specification of the Stream benchmark consists of three $N$ element vectors ${\bf A}, {\bf B}, {\bf B} : \mathbb{R}^N$ such that the total memory occupied is a significant fraction of the processor memory.  A specific series of operations is performed on these vectors in the following order:
\begin{eqnarray*}
   {\rm Copy:}   && ~~~~ {\bf C} = {\bf A} \\
   {\rm Scale:}  && ~~~~ {\bf B} = q ~ {\bf C} \\
   {\rm Add:}    && ~~~~ {\bf C} = {\bf A} + {\bf B} \\
   {\rm Triad:} && ~~~~ {\bf A} = {\bf B} + q ~ {\bf C} \nonumber
\end{eqnarray*}
The above operations are repeated $N_t$ times. The basic algorithm for Stream is shown in Algorithm~\ref{alg:SerialStream} (see caption for details).

\begin{algorithm}[Serial Stream]
The algorithm creates three $N$ element vectors and performs a series of operations on those vectors.  Each operation is timed using the \proc{tic} and \proc{toc} commands.
\label{alg:SerialStream}
\begin{codebox}
\Procname{$\proc{Stream}({\bf A}, {\bf B}, {\bf C} : \mathbb{R}^N, q : \mathbb{R},n : \mathbb{Z})$}
\li     $t_{copy}$, $t_{scale}$, $t_{add}$, $t_{copy}: \mathbb{R}$
\li     \For $i=1:N_t$
\li       \Do
\li           \proc{tic}
\li           ~~ ${\bf C} = {\bf A}$
\li           $t_{copy}$ += \proc{toc}
\li           \proc{tic}
\li           ~~ {\bf B} = q  {\bf C}
\li           $t_{scale}$ += \proc{toc}
\li           \proc{tic}
\li           ~~ {\bf C} = {\bf A} + {\bf B}
\li           $t_{add}$ += \proc{toc}
\li           \proc{tic}
\li           ~~ {\bf A} = {\bf B} + q  {\bf C}
\li           $t_{triad}$ += \proc{toc}
          \End
\end{codebox}
\end{algorithm}

The goal of Stream is to measure memory bandwidth in bytes per second. The benchmark requires the vectors to be 8-byte double-precision floating point values resulting in the following bandwidths formulas:
\begin{eqnarray*}
   {\rm Copy~Bandwidth:}   && ~~~~ 16 ~ N_t ~ N / t_{copy} \\
   {\rm Scale~Bandwidth:}  && ~~~~ 16 ~ N_t ~ N / t_{scale} \\
   {\rm Add~Bandwidth:}    && ~~~~ 24 ~ N_t ~ N / t_{add} \\
   {\rm Triad~Bandwidth:} && ~~~~ 24 ~N_t ~ N / t_{triad} \nonumber
\end{eqnarray*}
Validation is a critical part of any benchmark program. If the values of the vector ${\bf A}$ are initialized to $A_0$, then the final results can be checked against the formulas
\begin{eqnarray*}
   A_{N_t-1} & = & (2 q + q^2)^{N_t-1} ~ A_0 \\
   {\bf A}_{N_t}(:) & = & (2 q + q^2)^{N_t}  ~ A_0\\
   {\bf B}_{N_t}(:) & = & q ~ a_{N_t-1} \\
   {\bf C}_{N_t}(:) & = & (1+q) ~ A_{N_t-1} \nonumber
\end{eqnarray*}
Furthermore, selecting $q = \sqrt{2} -1$ so that $2 q + q^2 = 1$ ensures the values remain modest for large values of $N_t$.

\begin{figure}[]
\centerline{\includegraphics[width=1.0\columnwidth]{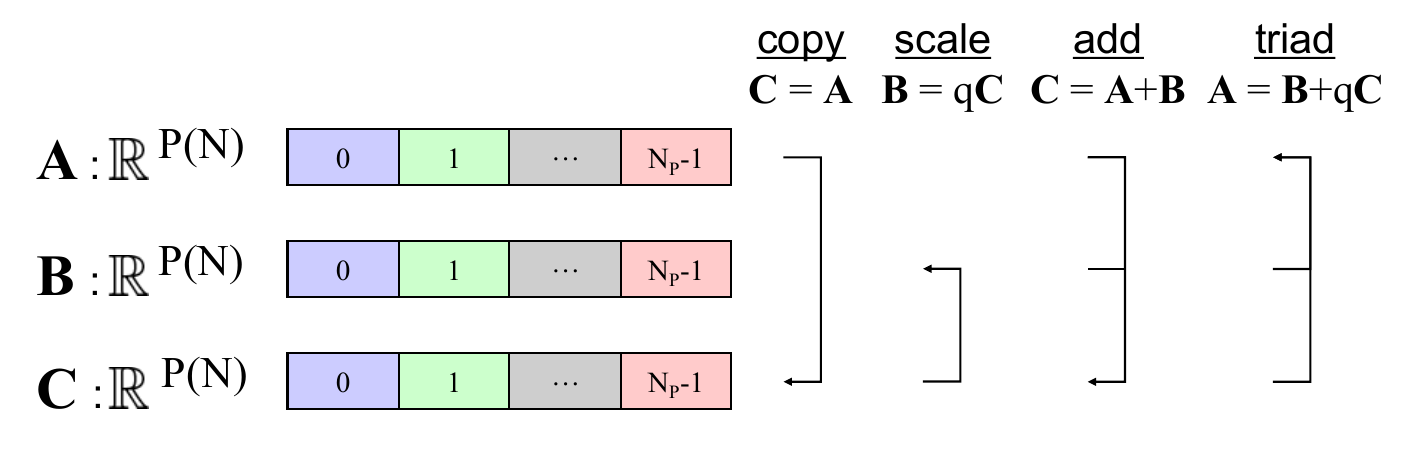}}
\caption{{\bf Parallel Stream Design.}
Each vector is a distributed array.  If each vector has the same parallel map, then the resulting program will require no communication.}
\label{fig:StreamDesign}
\end{figure}

Although Stream is very simple, it contains the essence of many programs.  Specifically, it operates on a series of vectors, with defined initial values.  It repeats a sequence of standard vector operations and produces answers that can be checked for correctness.

The parallel Stream program can be implemented quite easily using distributed arrays.  As long as all the vectors use the same distributed mapping, the resulting program will require no interprocessor communication.  The no communication requirement can be strictly enforced by making the program use the $.loc$ construct which specifies that only the part of the array that is local to the $P_{ID}$ is used.  This approach is generally recommended as it guarantees that no communication will take place.  If a mapping that required communication were accidentally used, then the resulting program will either produce an error or will fail to validate. This mapping results in the parallel algorithm shown using P notation in Algorithm~\ref{alg:ParallelStream}, which specifies which dimensions of the array are to be divided among different $P_{ID}$s.   A visual representation of the design for the parallel Stream algorithm is shown in Figure~\ref{fig:StreamDesign}.

\begin{algorithm}[Parallel Stream]
The parallel algorithm creates three $N$ element distributed vectors and performs a series of operations on just the local part of those vectors.  The resulting times can be averaged to obtain overall parallel bandwidths.
\label{alg:ParallelStream}
\begin{codebox}
\Procname{$\proc{ParallelStream}({\bf A}, {\bf B}, {\bf C} : \mathbb{R}^{P(N)}, q : \mathbb{R},n : \mathbb{Z})$}
\li     $t_{copy}$, $t_{scale}$, $t_{add}$, $t_{copy}: \mathbb{R}$
\li     \For $i=1:N_t$
\li       \Do
\li           \proc{tic}
\li           ~~ ${\bf C}.loc = {\bf A}.loc$
\li           $t_{copy}$ += \proc{toc}
\li           \proc{tic}
\li           ~~ {\bf B}.loc = q {\bf C}.loc
\li           $t_{scale}$ += \proc{toc}
\li           \proc{tic}
\li           ~~ {\bf C}.loc = {\bf A}.loc + {\bf B}.loc
\li           $t_{add}$ += \proc{toc}
\li           \proc{tic}
\li           ~~ {\bf A}.loc = {\bf B}.loc + q  {\bf C}.loc
\li           $t_{triad}$ += \proc{toc}
          \End
\end{codebox}
\end{algorithm}

\begin{CodeListing}[{\bf Parallel Stream Matlab program}]
\label{code:ParallelStreamMatlab}
\begin{codebox}
\\
\li 
\li 
\li
\li {\tt\footnotesize Nt = 10; N = Np*(2.$^\wedge$30);}
\li {\tt\footnotesize A0 = 1.0; B0 =2.0; C0 = 0.0;  q = sqrt(2)-1;}
\li {\tt\footnotesize     ABCmap = map([1 Np],\{\},0:Np-1);}  \>\>\>\>\>\>\>\>\>\> {\tt\footnotesize \%  Map.}
\li {\tt\footnotesize \% Allocate and initialize distributed vectors.}
\li {\tt\footnotesize     Aloc = {\color{blue} gpuArray}(local(zeros(1,N,ABCmap)))+A0;}
\li {\tt\footnotesize     Bloc = {\color{blue} gpuArray}(local(zeros(1,N,ABCmap)))+B0;}
\li {\tt\footnotesize     Cloc = {\color{blue} gpuArray}(local(zeros(1,N,ABCmap)))+C0;}
\li {\tt\footnotesize \% Initialize timers.}
\li {\tt\footnotesize TsumCopy=0.0; TsumScale=0.0;}
\li {\tt\footnotesize TsumAdd=0.0; TsumTriad=0.0;}
\li {\tt\footnotesize for i = 1:Nt} \>\>\>\>\>\>\>\>\>\> {\tt\footnotesize \% Run benchmark.}
\li \> {\tt\footnotesize     tic;} 
\li \> \> {\tt\footnotesize         Cloc(:,:) = Aloc;} \>\>\>\>\>\>\>\> {\tt\footnotesize \% Copy.}
\li \> {\tt\footnotesize     TsumCopy = TsumCopy + toc;}
\li \> {\tt\footnotesize     tic;}
\li \> \> {\tt\footnotesize         Bloc(:,:) = q*Cloc;} \>\>\>\>\>\>\>\> {\tt\footnotesize \% Scale.}
\li \> {\tt\footnotesize     TsumScale = TsumScale + toc;}
\li \> {\tt\footnotesize     tic;}
\li \> \> {\tt\footnotesize         Cloc(:,:) = Aloc + Bloc;} \>\>\>\>\>\>\>\> {\tt\footnotesize \% Add.}
\li \> {\tt\footnotesize     TsumAdd = TsumAdd + toc;}
\li \> {\tt\footnotesize     tic;}
\li \> \> {\tt\footnotesize         Aloc(:,:) = Bloc + q*Cloc;} \>\>\>\>\>\>\>\> {\tt\footnotesize \% Triad.}
\li \> {\tt\footnotesize     TsumTriad = TsumTriad + toc;}
\li {\tt\footnotesize end}
\end{codebox}
\end{CodeListing}

\section{Code}
  Code excerpts for distributed array Stream implementations in Matlab/Octave and Python are shown in Code Listings \ref{code:ParallelStreamMatlab} and \ref{code:ParallelStreamPython}.  These codes closely mirror the parallel Stream algorithm. The parallel code can be implemented by simply converting the vectors from regular arrays to distributed arrays.  This same code is run by every instance with the only difference between these instances is that each has a unique $P_{ID}$ that is used to determine which parts of a distributed array belong to each instance (i.e., ${\bf A}.loc$, ${\bf B}.loc$, and ${\bf C}.loc$).

The descriptions of Code Listings \ref{code:ParallelStreamMatlab} and \ref{code:ParallelStreamPython} are as follows:

\noindent {\bf Lines 1--3} of the Python code import the required Python packages.  The CuPy package highlighted in blue enables running on GPUs.

\noindent {\bf Line 6} creates the parallel map {\tt ABCmap} specifying that the $1{\times}N$ element row vectors are to have their columns divided equally among all {\tt Np} over processes with $P_{ID}$ = {\tt 0}, ..., {\tt Np-1}.

\noindent {\bf Lines 8--10} create the distributed vectors {\tt A}, {\tt B}, and {\tt C}; extract their local parts; and then initialize them to {\tt A0}, {\tt B0}, and {\tt C0}.  Because the program requires no communication, the distributed arrays {\tt A}, {\tt B}, and {\tt C} are never actually allocated.  Instead, only their local parts {\tt Aloc}, {\tt Bloc}, and {\tt Cloc} are  created. This simplifies the program and reduces the total memory required to run the program by eliminating unnecessary duplication. The Matlab Parallel Computing Toolbox (PCT) function {\tt \color{blue} gpuArray} (currently not implemented in

\begin{CodeListing}[{\bf Parallel Stream Python program}]
\label{code:ParallelStreamPython}
\begin{codebox}
\\
\li {\tt\footnotesize import numpy as np}
\li {\tt\footnotesize {\color{blue} import cupy as cp}}
\li {\tt\footnotesize import pPython as GPC}
\li {\tt\footnotesize Nt = 10; N = GPC.Np*(2**30)}
\li {\tt\footnotesize A0 = 1.0; B0 =2.0; C0 = 0.0;  q = np.sqrt(2)-1}
\li {\tt\footnotesize     ABCmap = Dmap([1,Np],\{\},range(Np))}  \>\>\>\>\>\>\>\>\>\>\> {\tt\footnotesize \#  Map.}
\li {\tt\footnotesize \# Allocate and initialize distributed vectors.}
\li {\tt\footnotesize     Aloc = {\color{blue} cp.array}(local(zeros(1,N,map=ABCmap)))+A0}
\li {\tt\footnotesize     Bloc = {\color{blue} cp.array}(local(zeros(1,N,map=ABCmap)))+B0}
\li {\tt\footnotesize     Cloc = {\color{blue} cp.array}(local(zeros(1,N,map=ABCmap)))+C0}
\li {\tt\footnotesize \# Initialize timers.}
\li {\tt\footnotesize TsumCopy=0.0; TsumScale=0.0}
\li {\tt\footnotesize TsumAdd=0.0; TsumTriad=0.0}
\li {\tt\footnotesize for i in range(Nt):} \>\>\>\>\>\>\>\>\>\> {\tt\footnotesize \# Run benchmark.}
\li \> {\tt\footnotesize     tic} 
\li \> \> {\tt\footnotesize         Cloc[:,:] = Aloc} \>\>\>\>\>\>\>\> {\tt\footnotesize \# Copy.}
\li \> {\tt\footnotesize     TsumCopy += timer()-tic}
\li \> {\tt\footnotesize     tic}
\li \> \> {\tt\footnotesize         Bloc[:,:] = q*Cloc} \>\>\>\>\>\>\>\> {\tt\footnotesize \# Scale.}
\li \> {\tt\footnotesize     TsumScale += timer()-tic}
\li \> {\tt\footnotesize     tic}
\li \> \> {\tt\footnotesize         Cloc[:,:] = Aloc + Bloc} \>\>\>\>\>\>\>\> {\tt\footnotesize \# Add.}
\li \> {\tt\footnotesize     TsumAdd += timer()-tic}
\li \> {\tt\footnotesize     tic}
\li \> \> {\tt\footnotesize         Aloc[:,:] = Bloc + q*Cloc} \>\>\>\>\>\>\>\> {\tt\footnotesize \# Triad.}
\li \> {\tt\footnotesize     TsumTriad += timer()-tic}
\li {\tt\footnotesize  \# End loop.}
\end{codebox}
\end{CodeListing}

\noindent  Octave) and the Python CuPy {\tt \color{blue} cp.array} function copy  {\tt Aloc}, {\tt Bloc}, and {\tt Cloc} to the GPU memory; subsequent operations on these variables will be automatically performed on the GPU.

\noindent {\bf Lines 14--26} execute the copy, scale, add, and triad operations on {\tt Aloc}, {\tt Bloc}, and {\tt Cloc}.  The time for each of these operations is summed for each trial.  For GPUs a PCT {\tt \color{blue} wait}  command or CuPy {\tt \color{blue} synchronize} command may need to be called prior to each {\tt tic} or {\tt toc} command to ensure the operations are completed on the GPU.

The above code is a good example of the distributed array programming approach.  There are a number of properties of the distributed array approach that can be observed in this code example
\begin{itemize}
\item Low Code Impact.  The program has been converted from a serial program to a parallel program with very few additional lines of code.
\item Small parallel library footprint. Only two parallel library functions were required: {\tt map} and {\tt local}.
\item Scalable.  The code can be run on any problem size or number of processors such that $N > N_P$, provided the vectors all have identical maps.
\item Bounded communication.  Because local variables are used, there is strict control on when communication takes place.
\item Map independence.  As long as the same map is used for all three vectors, the program will work for any distribution in the second dimension (i.e., block, cyclic, or block-cyclic).  To make the program map independent for any combination of maps would require replacing the local operations with global operations. For example, if copy was implemented using {\tt C(:,:) = A}, then it would run correctly regardless of the map.  However, if {\tt A} and {\tt C} had different maps, then significant communication would be required.
\item Performance guarantee.  Because the copy, scale, add, and triad operations commands are working local variables that are  regular Matlab/Octave/Python numeric arrays, there is a guarantee that there is no hidden performance penalty when running these lines of code.
\end{itemize}

\section{Computer Hardware}

Our team has developed a high-productivity scalable platform---the MIT SuperCloud---for providing scientists and engineers the tools they need to analyze large-scale dynamic data \cite{kepner2012dynamic, gadepally2018hyperscaling, 8547629}.  The MIT SuperCloud provides interactive analysis capabilities  accessible from high level programming environments (Python, Julia, Matlab/Octave) that scale to thousands of processing nodes.  MIT SuperCloud maintains a diverse set of hardware running an identical software stack that allows direct comparison of hardware from different eras.

\begin{table}
\caption{Computer Hardware Specifications}
\vspace{-0.25cm}
MIT SuperCloud maintains a diverse set of hardware running an identical modern software stack providing a unique platform for comparing performance over different eras. GPUs are listed below their host systems. IBM Blue Gene P (bg-p) system was hosted at Argonne National Laboratory.
\begin{center}
\includegraphics[width=\columnwidth]{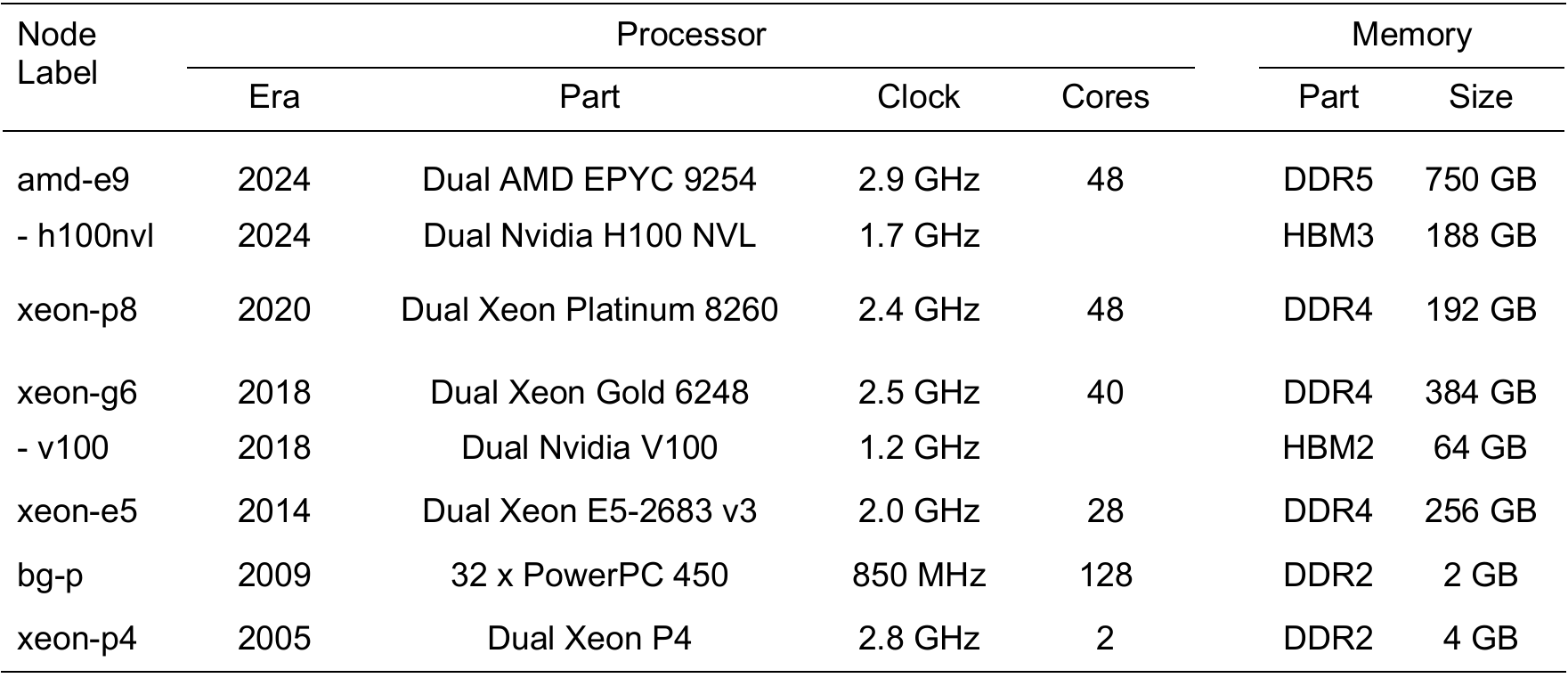}
\end{center}
\label{tab:HardwareTable}
\end{table}%

A typical benchmarking run can be launched in a few seconds using the MIT SuperCloud triples-mode hierarchical launching system \cite{8547629} enabling rapid interactive benchmarking.  The launch parameters were [$N_{node}$ $N_{ppn}$ $N_{tpn}$], corresponding to $N_{node}$ nodes, $N_{ppn}$ Matlab/Octave or Python processes per node, and $N_{tpn}$ OpenMP threads per process.   The total number of processes is given by $N_P = N_{node} N_{ppn}$. On each node, each of the $N_{ppn}$ processes and their corresponding $N_{tpn}$ threads were pinned to adjacent cores to minimize interprocess contention and maximize cache locality for the Stream OpenMP threads \cite{byun2019optimizing}.  Within each Matlab/Octave or Python process, the OpenMP parallelism is used as provided by their math libraries.  At the end of the processing the results were aggregated using asynchronous file-based messaging \cite{byun2019large}. Triples mode makes it easy to explore horizontal scaling  across nodes, vertical scaling by examining combinations of processes and threads on a node, and temporal scaling by running on diverse hardware from different eras.

The computing hardware consists of many different types of nodes used over two decades (see Table~\ref{tab:HardwareTable}).  With the exception of the bg-p nodes all are multicore x86 compatible.  The MIT SuperCloud maintains the same modern software across all nodes, which allows for direct comparison of hardware performance differences.  In addition, results from prior distributed array Stream benchmarking are included \cite{haney2004pmatlab, byun2010toward}.  The IBM Blue Gene P (bg-p) results were collected on the Surveyor system hosted at Argonne National Laboratory.

\begin{table}
\caption{Single Node STREAM Parameteres}
\vspace{-0.25cm}
Number of trials ($N_t$) and vector size per process instance $N/N_P$ for different hardware.  STREAM benchmark parameters were chosen to balance consistency with the memory capacity of the different hardware.  For multiple nodes the parameters highlighted in {\bf bold} were used.
\begin{center}
\includegraphics[width=\columnwidth]{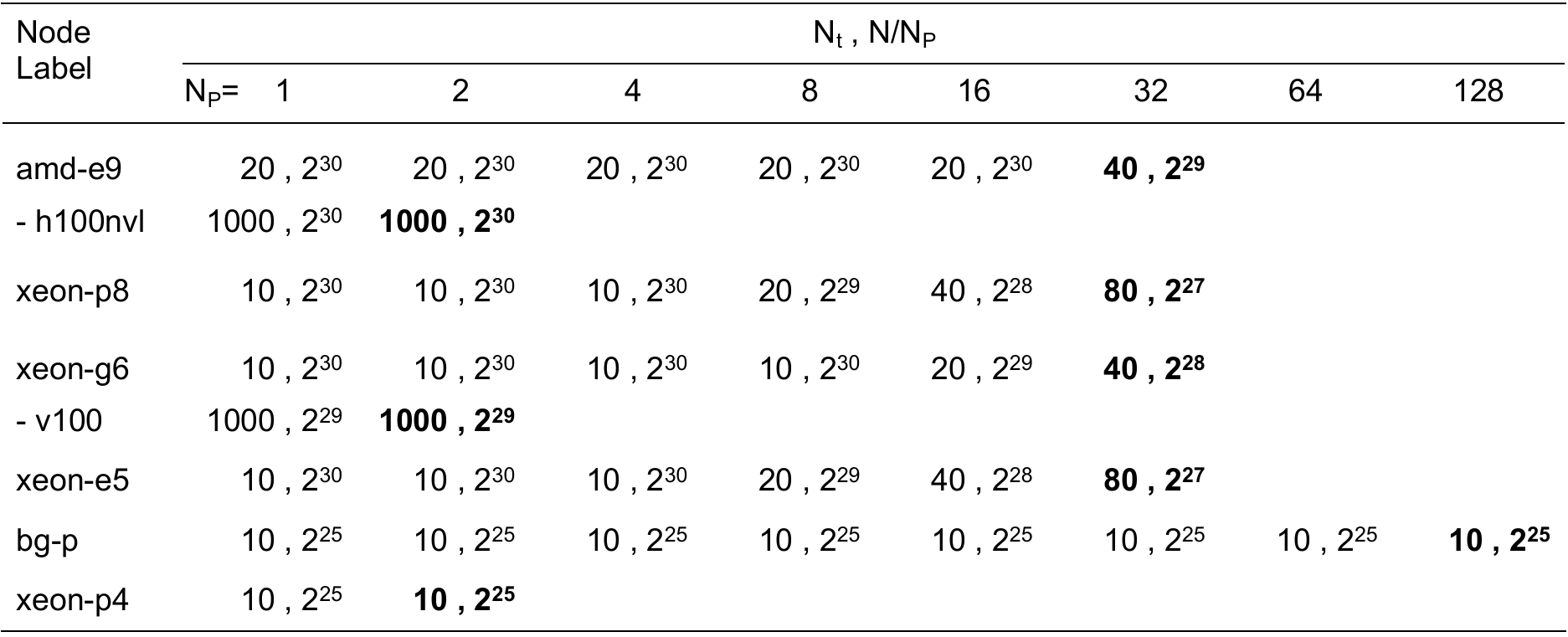}
\end{center}
\label{tab:STREAMparameters}
\end{table}%

Matlab, Octave, and Python Stream distributed array implementations for the different hardware configurations (see Table~\ref{tab:HardwareTable}) were run with the parameters listed in Table~\ref{tab:STREAMparameters}.  The Stream vector size parameter $N$ was chosen with a base value of $N=2^{30}$ which is the maximum PCT and CuPy currently allow for GPU calculations.  Within a node $N$ was then scaled with $N_P$ (i.e., constant $N/N_P$) to the maximum memory allowed on the node.  Once the node memory maximum was hit, $N$ was kept constant  (i.e., decreasing $N/N_P$) and $N_t$ was increased to keep the relative run time around a few hundred seconds.   For multiple nodes, $N$ was again scaled with $N_P$ (i.e., constant $N/N_P$) and $N_t$ was held fixed.

\begin{figure*}[]
\centering
\includegraphics[width=0.9\columnwidth]{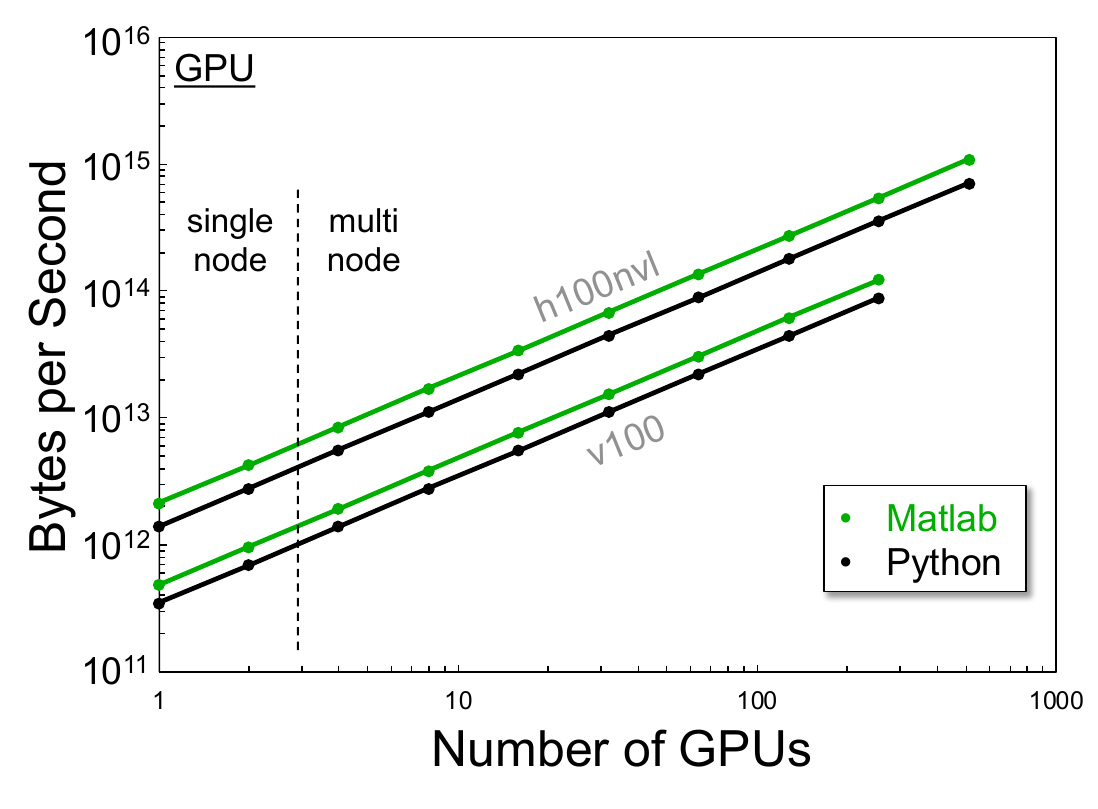}
\\
\includegraphics[width=0.9\columnwidth]{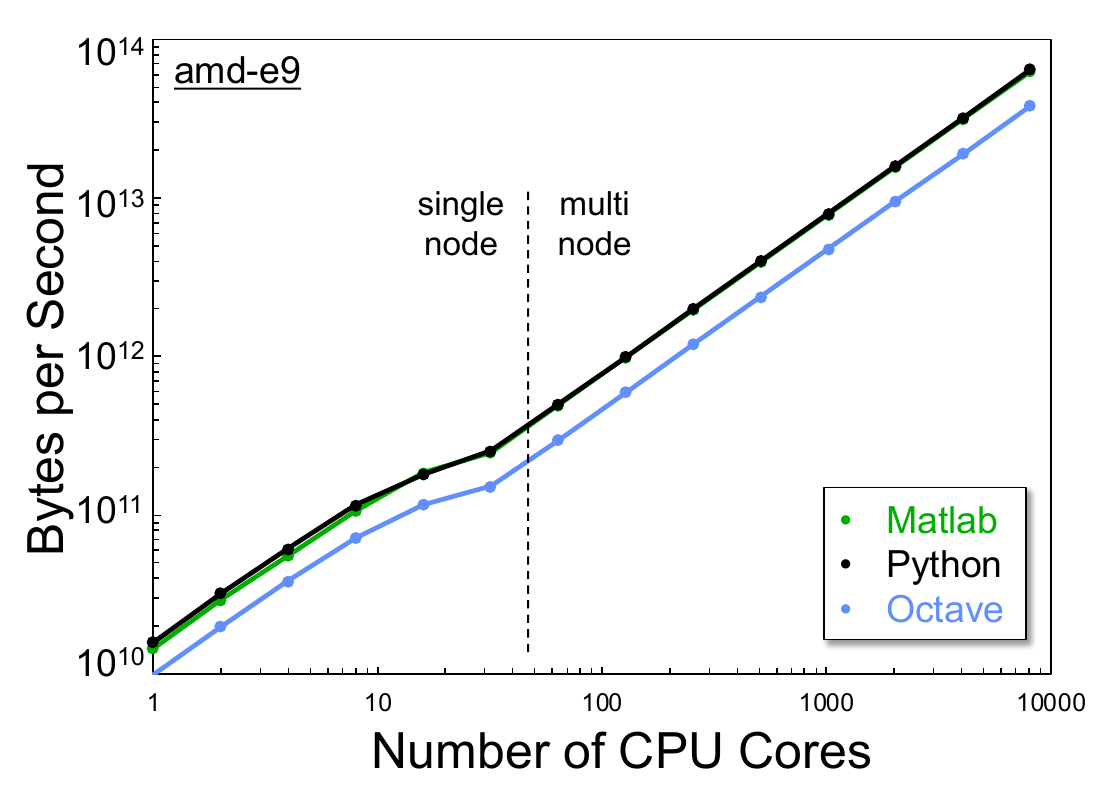}
\includegraphics[width=0.9\columnwidth]{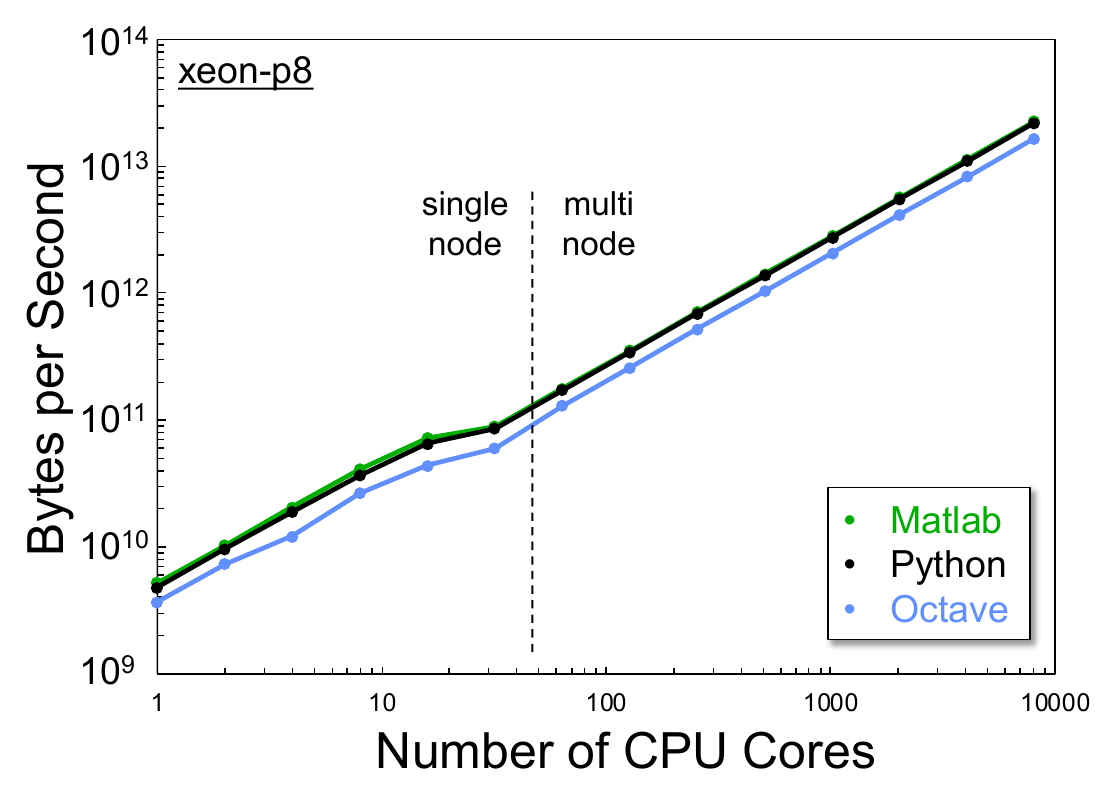}
\includegraphics[width=0.9\columnwidth]{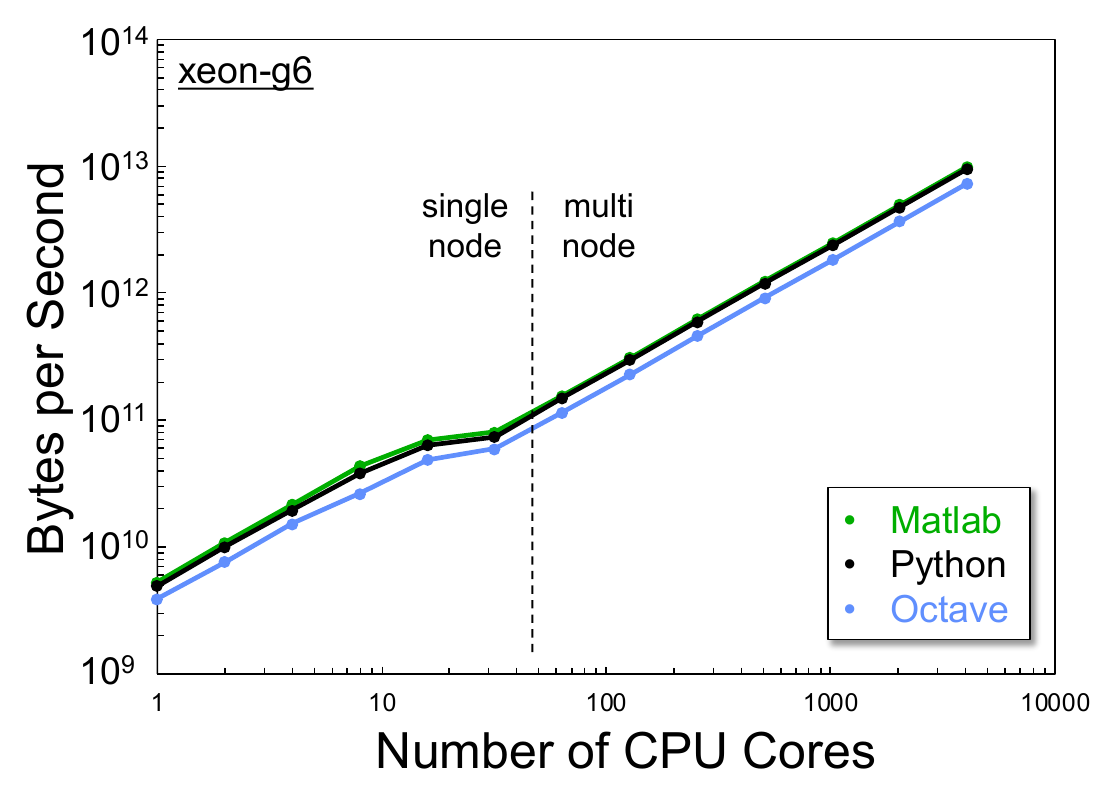}
\includegraphics[width=0.9\columnwidth]{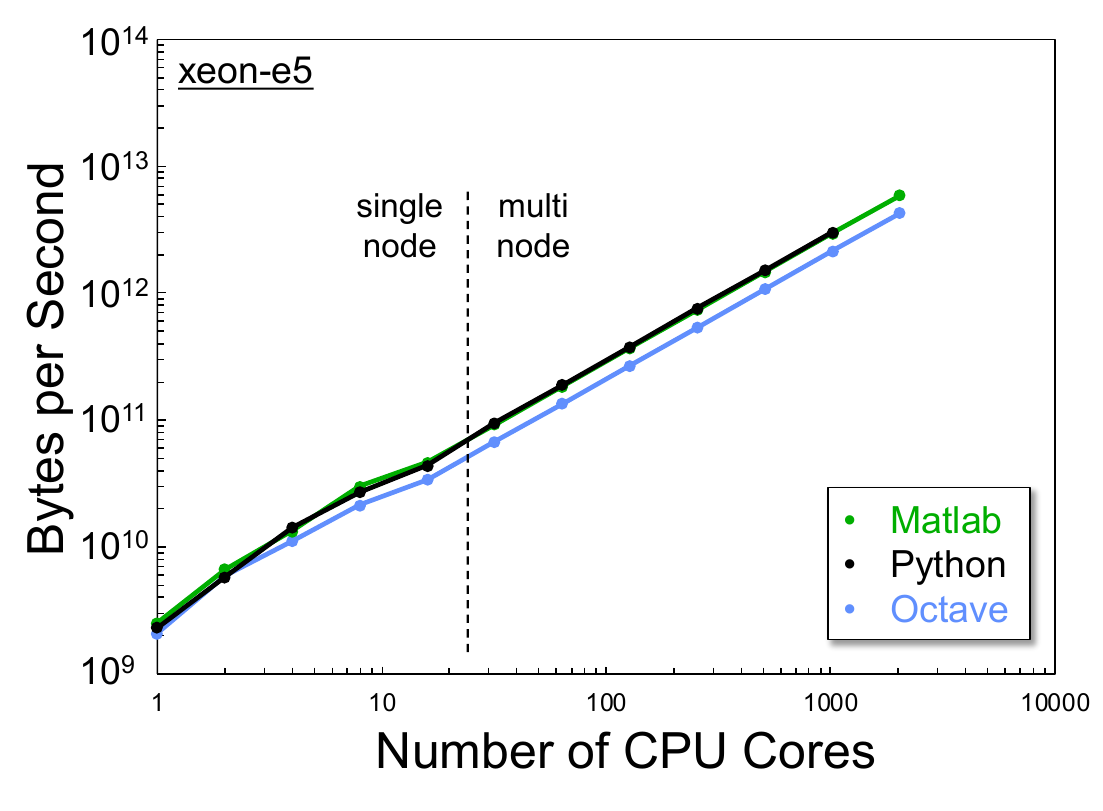}
\includegraphics[width=0.9\columnwidth]{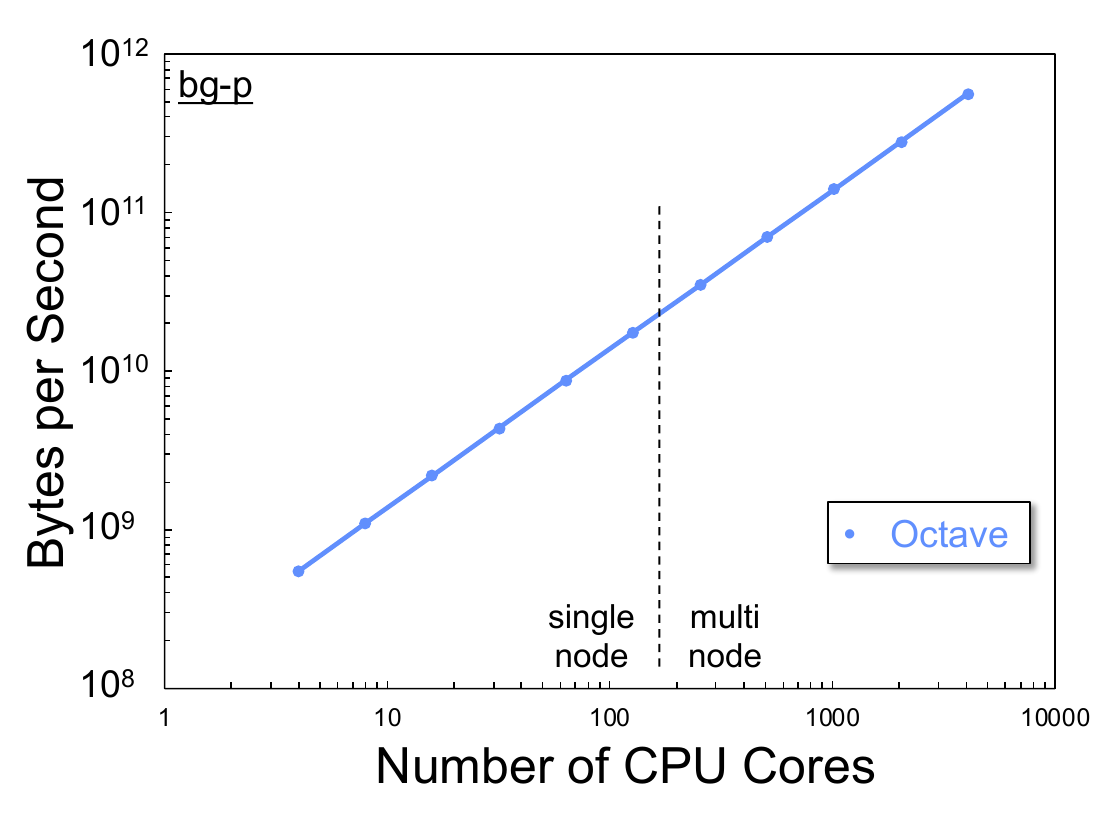}
\includegraphics[width=0.9\columnwidth]{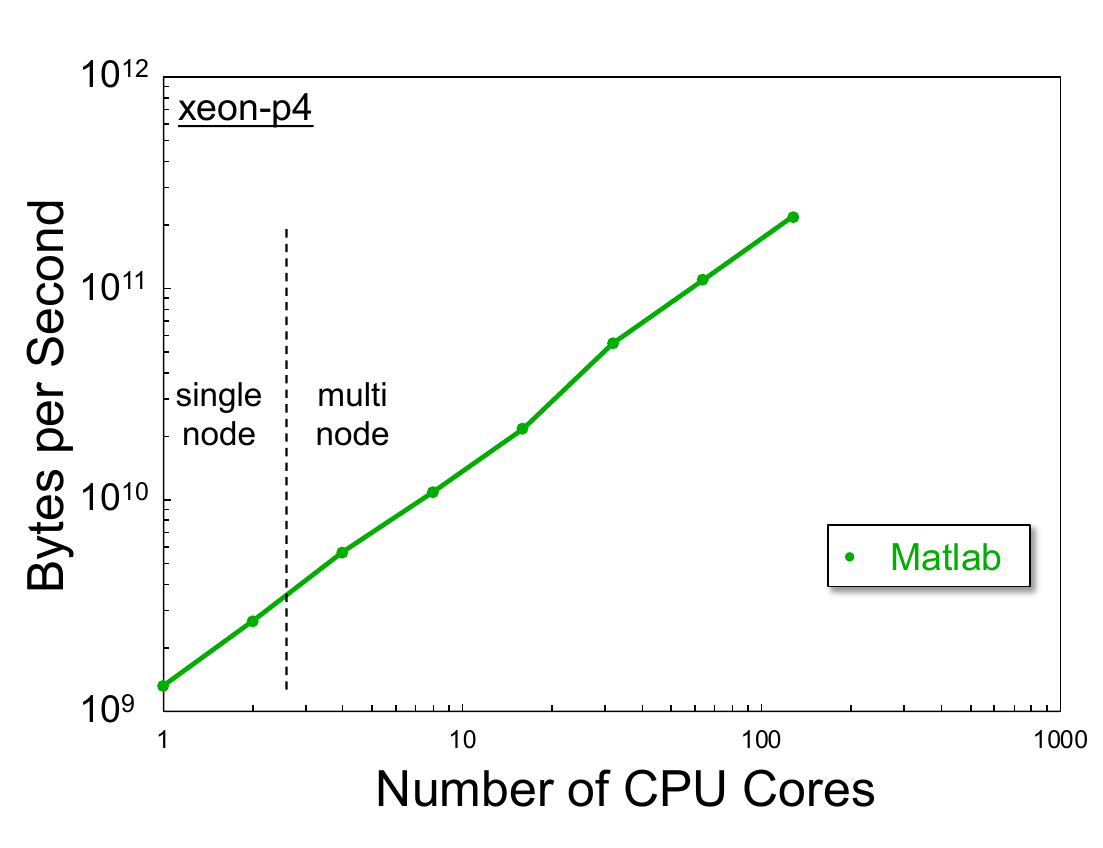}
\caption{{\bf Measured Bandwidth}. Matlab, Octave, and Python Stream triad bandwidth using distributed arrays for the different hardware configurations (see Table~\ref{tab:HardwareTable}) run with the parameters listed in Table~\ref{tab:STREAMparameters}.  The bg-p data was adapted from \cite{byun2010toward}.  The xeon-p4 data was adapted from \cite{haney2004pmatlab}.    All plots show excellent vertical scaling within a node, horizontal scaling across nodes, and temporal scaling over multiple eras of hardware.}
\label{fig:MeasuredBandwidth}
\end{figure*}

\section{Performance Results}

Figure~\ref{fig:MeasuredBandwidth} shows the within node (vertical scaling) and across node (horizontal scaling) Stream triad bandwidth for all the different configurations of hardware listed in Table~\ref{tab:HardwareTable}.  The \mbox{bg-p} data was adapted from \cite{byun2010toward}.  The xeon-p4 data was adapted from \cite{haney2004pmatlab}.    All plots show excellent vertical scaling within a node, horizontal scaling across nodes, and temporal scaling over multiple eras of hardware.   Stream copy, add, and scale following a similar pattern, but only the triad results are shown.  The Octave interpreter defers the first copy in the Stream benchmark and folds it into triad, which is why the Octave results are generally $\sim30\%$ lower.

\begin{figure}[]
\centering
\includegraphics[width=\columnwidth]{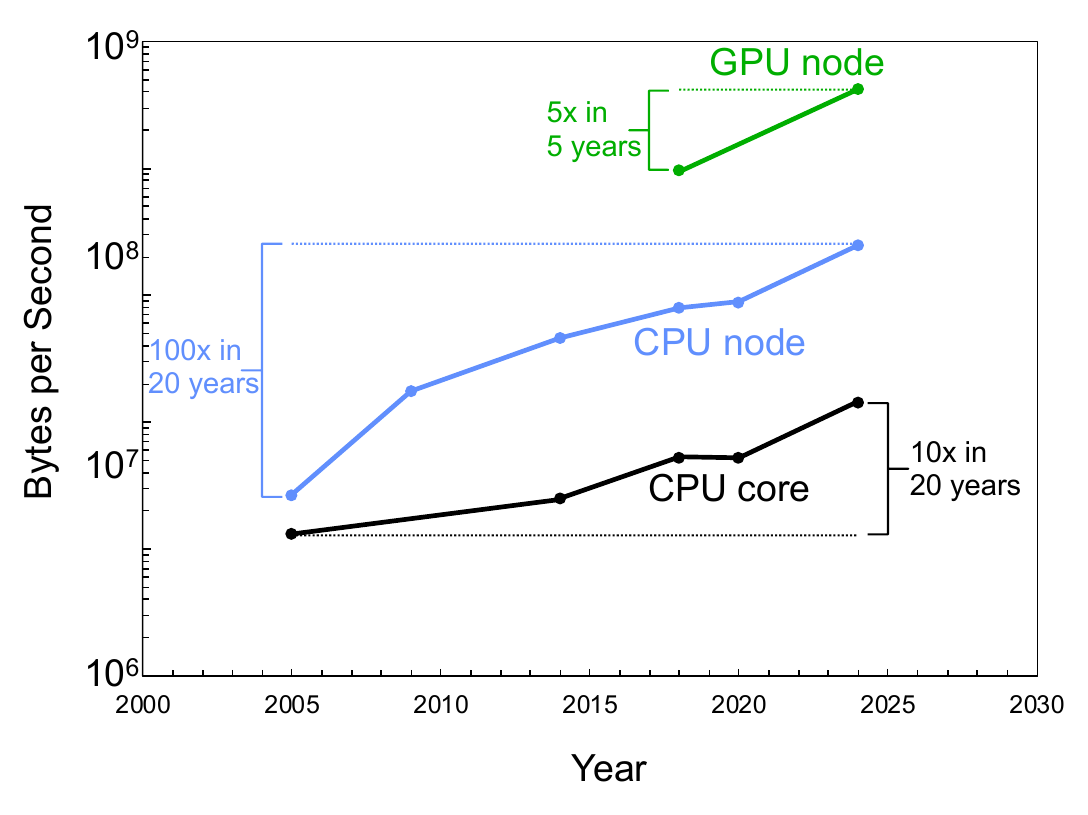}
\caption{{\bf Temporal Scaling}.  Stream triad bandwidth of hardware at different eras for a single process on a single core running a single thread (bottom black line), multiple processes on multiple cores each running a single thread (middle blue line), and 2 processes on a 2 GPU node (top green line).  These benchmark data indicate a 10x increase in single-core bandwidth over 20 years, a 100x increase in single node bandwidth over 20 years, and a 5x increase in single GPU node performance over 5 years.
}
\label{fig:SingleNode-Year}
\end{figure}

Temporal benchmarking compares the performance of computing hardware from different eras.  The MIT SuperCloud provides a unique ability to directly compare hardware from different eras using exactly the same modern software stack.  Figure~\ref{fig:SingleNode-Year} shows the best single-core and single-node bandwidth from Figure~\ref{fig:MeasuredBandwidth} and plots the values versus the computing hardware era from Table~\ref{tab:HardwareTable}. These benchmark data indicate a 10x increase in single-core bandwidth over 20 years, a 100x increase in single node bandwidth over 20 years, and a 5x increase in single GPU node performance over 5 years.

\section{Conclusion}

 A wide range of applications are currently being enabled by high level programming languages and GPU accelerators.  Retaining productivity requires effective abstractions for achieving performance within a compute node (vertical scaling), across compute nodes (horizontal scaling), and over different generations of hardware (temporal scaling).   One such abstraction that enables high level programming to achieve highly scalable performance are distributed arrays.  By deriving parallelism from data locality, which naturally leads to high memory bandwidth efficiency, distributed arrays provide a straightforward approach for achieving parallel performance.  Using the STREAM memory bandwidth benchmark distributed array performance has been explored on a variety of hardware.  Linear scaling across nodes was demonstrated.  Direct comparison of hardware improvements for memory bandwidth over this time range show a 10x increase in CPU core bandwidth over 20 years, a 100x increase in CPU node bandwidth over 20 years, and a 5x increase in GPU node bandwidth over 5 years.  Running on hundreds of MIT SuperCloud nodes simultaneously achieved a sustained bandwidth $>$1 PB/s.    Future work will seek to extend the distributed array programming model to other high level languages and other hardware platforms.
\section*{Acknowledgement}

The authors wish to acknowledge the following individuals for their contributions and support:  Bob Bond, Alan Edelman, Peter Fisher, Jeff Gottschalk, Chris Hill, Charles Leiserson, Kristen Malvey, Sandy Pentland, Heidi Perry, Sandeep Pisharody,  Steve Rejto,  Mark Sherman, Marc Zissman.




\bibliographystyle{ieeetr}

\bibliography{EasyAcceleration}
%

\end{document}